\documentclass[final,3p,times,twocolumn]{elsarticle}

\usepackage{amssymb,amsmath,amsthm,bm}
\usepackage{amsopn}
\usepackage{geometry}
\usepackage{algorithm}
\usepackage{algorithmic}
\usepackage{graphicx}

\usepackage[retainorgcmds]{IEEEtrantools}
\graphicspath{{figures/}}
\newlength\imagewidth
\newtheorem{Property}{Property}

\journal{Signal Processing}
\usepackage{color}
\definecolor{dgreen}{rgb}{0,.6,0}

\usepackage{lineno}

\begin{document}

\begin{frontmatter}

\title{Breaking a novel colour image encryption algorithm based on chaos}

\author[cn-xtu-cie,cn-xtu-icip]{Chengqing Li\corref{corr}}
\ead{chengqingg@gmail.com}

\author[cn-xtu-cmc]{Yu Zhang}

\author[cn-xtu-cie]{Rong Ou}

\author[hk-cityu]{Kwok-Wo Wong}

\cortext[corr]{Corresponding author.}

\address[cn-xtu-cie]{College of Information Engineering,
Xiangtan University, Xiangtan 411105, Hunan, China}

\address[cn-xtu-icip]{MOE (Ministry of Education) Key Laboratory of Intelligent Computing and Information Processing, Xiangtan University, China}

\address[cn-xtu-cmc]{School of Mathematics and Computational Science,
Xiangtan University, Xiangtan 411105, Hunan, China}


\address[hk-cityu]{Department of Electronic Engineering, City University of Hong Kong, Hong Kong, China}

\begin{abstract}
Recently, a colour image encryption algorithm based on chaos was proposed by cascading two position permutation operations and one substitution operation,
which are all determined by some pseudo-random number sequences generated by iterating the Logistic map. This paper evaluates the security level of the encryption algorithm and finds that the position permutation-only part and the substitution part can be separately broken with only $\lceil (\log_2(3MN))/8 \rceil$ and $2$ chosen plain-images, respectively, where $MN$ is the size of the plain-image. Concise theoretical analyses are provided to support the chosen-plaintext attack, which are verified by experimental results also.
\end{abstract}


\end{frontmatter}

\section{Introduction}
\label{sec:intro}

Security of multimedia data (image, video, audio/speech) become more and more important as it is transmitted over all kinds of wired/wireless networks
more and more frequently. Both design and security analysis of multimedia encryption algorithms have been received keen attention of the related
researchers in the past decade \cite{MinWu:joint:TIP06,Goce:cryptanalysis:TM08,Zhou:AttackArithmeticCoding:TSP09,Liaoxf:wave:SP10,Uhl:SurveyAVC:IEEETCASVT12}. Due to the subtle similarity between some dynamical properties of chaos,
like sensitivity to changes of initial condition and control parameter of chaotic systems, and the basic properties of cryptography, diffusion and confusion, chaos was considered as a special way to design secure and efficient encryption algorithm \cite{YaobinMao:CSF2004,Tong:compound:SP09,ChenJY:Joint:TCSII11}. As image data is a representative form of multimedia data, and it helps to show the claimed good performances of the proposed encryption algorithms, most chaos-based encryption algorithms adopt image data as encryption object.

According to the record of \textit{Web of Science}, more than four hundred papers on designing chaos-based image encryption schemes were published between 1997 and 2011 (inclusive). Meanwhile, no more than one hundred and half papers on security analysis of chaos-based image encryption schemes were published. Short of scrutiny on the security
makes many chaos-based image encryption schemes are insecure against some conventional attacks, such as known/chosen-plaintext attack and chosen-ciphertext attack \cite{David:AttackingChaos08,Li:BreakImageCipher:IVC09,Li:AttackingIVC2009,SolakErcan:AnaFridrich:BAC10}. Some representative chaos-based encryption algorithms and a general framework evaluating security of this class of encryption algorithms were concluded in \cite{AlvarezLi:Rules:IJBC2006}. In many chaos-based image encryption algorithms, a chaos system, composed of one or more chaotic maps, is used to generate pseudo-random number sequence (PRNS), which is then adopted to determine and control combination of some basic encryption functions \cite{Li:AttackHDSP2006,Seyed:piecewise:SP12}. In digital domain, finite precision computation and quantization process make some dynamical properties of chaos system be degenerated in some form, which may cause potential threat to security of the chaos-based encryption algorithms \cite{FChen:PeriodArnold:TIT12}.

The present paper analyzes the security of the image encryption algorithms proposed in \cite{WangXY:3operations:SP12} and finds that the three basic encryption operations of the algorithm
are all \textit{key-invertible}, i.e. the unknown information controlling an encryption operation can be derived directly from the input and its output result. Furthermore, the three encryption functions are run independently. So, the position permutation part and the substitution part of the image encryption algorithm under study can be broken separately with a few chosen plain-images. Both detailed theoretical analyses and experimental results are presented to support the chosen-plaintext attack.

The rest of this paper is organized as follows. The next section introduces the image encryption algorithm under study briefly. Section~\ref{sec:cpa} presents an efficient chosen-plaintext attack on the encryption algorithm with some experimental results. The last section concludes the paper.

\section{The colour image encryption algorithm under study}

The plaintext of the encryption algorithm under study is a RGB colour image of size $M\times N$ (height$\times$width), which can be represented as
a $M\times N\times 3$ matrix of pixel values $\bm{I}=\{I(i, j, k)\}_{i=0, j=0, k=0}^{M-1, N-1, 2}=\{(R(i, j), G(i, j), B(i, j))\}_{i=0, j=0}^{M-1, N-1}$. Similarly,
the corresponding cipher-image is denoted by $\bm{I}'=\{I'(i, j, k)\}_{i=0, j=0, k=0}^{M-1, N-1, 2}=\{(R'(i, j), G'(i, j), B'(i, j))\}_{i=0, j=0}^{M-1, N-1}$.
Then, the colour image encryption algorithm under study can be described as follows\footnote{To make the presentation more concise and complete, some notations
in the original paper \cite{WangXY:3operations:SP12} are modified under the condition that essential form of the encryption algorithm is kept unchanged.}.

\begin{itemize}
\item \textit{The secret key} is composed of two positive integers $m_1$, $m_2$, and two sets of initial condition and control parameter
of the logistic map
\begin{equation}
f(x)=\mu \cdot x \cdot (1-x),
\label{eq:logistic}
\end{equation}
$(x_0, \mu_0)$, $(x^*_0, \mu_0^*)$, where $x_0, x^*_0\in (0, 1)$, and $\mu_0, \mu_0^*\in (3.5699456, 4)$.

\item \textit{The initialization procedure}:

(1) Iterate the logistical map~(\ref{eq:logistic}) $m_1$ times from initial condition $x_0$ to obtain a new initial condition under fixed
control parameter $\mu_0$. Then, further iterate it $3M$ times to get a chaotic states sequence $\{X_l\}_{l=0}^{3M-1}$. Finally, a permutation
sequence $\{T_l\}_{l=0}^{3M-1}$ is derived by comparing $\{X_l\}_{l=0}^{3M-1}$ and its sorted version, where $X_{T_l}$ is the $l$-th largest element in the sequence $\{X_l\}_{l=0}^{3M-1}$.

(2) Iterate the logistical map~(\ref{eq:logistic}) $m_2$ times from initial condition $x_0^*$ to obtain a new initial condition under fixed
control parameter $\mu_0^*$. Then, further iterate it $3MN$ times to get a chaotic states sequence $\{X^*_l\}_{l=0}^{3MN-1}$. For $i=0\sim M-1$,
obtain another permutation sequence $\{T^*_{i,l}\}_{l=0}^{3N-1}$ by comparing $\{X^*_{3iN+l}\}_{l=0}^{3N-1}$ and its sorted version, where $X^*_{3iN+T^*_{i, l}}$ is the $l$-th largest elements in sequence $\{X^*_{3iN+l}\}_{l=0}^{3N-1}$.

(3) Generate a PRNS $\{Y_l\}_{l=0}^{3MN-1}$ from the sequence $\{X^*_l\}_{l=0}^{3MN-1}$ via
$Y_l=\lfloor X^*_l\cdot 10^{14} \rfloor \bmod 3$, where
\begin{equation*}
(a \bmod b)=a-b\cdot \lfloor a/b\rfloor
\end{equation*}
when $b\neq 0$.

(4) To make the numbers of the three different elements in $\{Y_l\}_{l=0}^{3MN-1}$ are all equal to $MN$,
update the last $3MN-1$ elements as follows: for $l=1\sim 3MN-1$, set
\begin{equation*}
Y_l=
\begin{cases}
1, & \text{if } Y_l=0,  n_0 \geq MN \text{ and }n_1<MN,   \\
2, & \text{if } Y_l=0,  n_0 \geq MN \text{ and }n_1\ge MN, \\
2, & \text{if } Y_l=1,  n_1 \geq MN \text{ and }n_2< MN,   \\
0, & \text{if } Y_l=1,  n_1 \geq MN \text{ and }n_2\ge MN, \\
0, & \text{if } Y_l=2,  n_2 \geq MN \text{ and }n_0<MN,    \\
1, & \text{if } Y_l=2,  n_2 \geq MN \text{ and }n_0\ge MN,
\end{cases}
\end{equation*}
where $n_0$, $n_1$, $n_2$ represent the number of $0$, $1$, $2$ in $\{Y_i\}_{i=0}^{l-1}$, respectively.

(5) Generate another PRNS $\{Z_l\}_{l=0}^{3MN-1}$ from the sequence $\{X^*_l\}_{l=0}^{3MN-1}$ via
$Z_l=\lfloor X^*_l\cdot 10^{14} \rfloor \bmod 256$.

\item \textit{The encryption procedure} is a simple concatenation of the following three encryption operations.

(1) \textit{Row permutation:}
for $i=0\sim M-1$, $j=0\sim N-1$, $k=0\sim 2$, set
\begin{equation*}
I^*(i, j, k)=I(i^*, j, k^*),
\end{equation*}
where $i^*=T_{kM+i}\bmod M, k^*=\lfloor T_{kM+i}/M\rfloor$.

(2) \textit{Column permutation:}
for $i=0\sim M-1$, $j=0\sim N-1$, $k=0\sim 2$, set
\begin{equation*}
I^{**}(i, j, k)=I^*(i, j^{**}, k^{**}),
\end{equation*}
where $j^{**}=T^*_{i, kN+j }\bmod N, k^{**}=\lfloor T^*_{i, kN+j}/N\rfloor$.

(3) \textit{Substitution:}
First, let
\begin{equation}
I'(0, 0, Y_0)=(I^{**}(0, 0, Y_0) + Z_0)\bmod 256.
\label{eq:substitutionitem}
\end{equation}
Then, one pixel is selected iteratively from the other un-encrypted pixels of the intermediate image $\bm{I}^{**}=\{I^{**}(i, j, k)\}_{i=0, j=0, k=0}^{M-1, N-1, 2}$ according to
a PRNS $\{Y_l\}_{l=1}^{3MN-1}$, determining which channel's pixel is chosen. The selected pixels are encrypted by
the previous selected pixel, the corresponding cipher-pixel and a pseudo-random number as follows:
calculate
\begin{multline}
I'(i, j, k) = (I^{**}(i, j, k) +I^{**}(i', j', k') \\+ I'(i', j', k')+ Z_l)\bmod 256
\label{eq:substitution}
\end{multline}
for $l=1\sim 3MN-1$, where
\begin{alignat*}{3}
i &=\lfloor n_{k}/N \rfloor,   & \quad j &=n_{k}\bmod N,  & \quad k & =Y_{l},\\
i'&=\lfloor n_{k'}/N \rfloor,  & \quad j'&=n_{k'}\bmod N, & \quad k' & =Y_{l-1},
\end{alignat*}
$n_{k}$ and $n_{k'}$ represent the number of $k$ and $k'$ in $\{Y_t\}_{t=0}^{l}$ and $\{Y_t\}_{t=0}^{l-1}$, respectively.

\item \textit{The decryption procedure} is similar to the encryption one except the following points:
(1) the above encryption operations are run in a reverse order; (2) the permutation sequences are replaced by their invertible versions;
(3) equation~(\ref{eq:substitutionitem}) and Eq.~(\ref{eq:substitution}) are replaced by
\begin{equation*}
I^{**}(0, 0, Y_0)=( I'(0, 0, Y_0)- Z_0)\bmod 256
\end{equation*}
and
\begin{multline*}
I^{**}(i, j, k) = (I'(i, j, k) - I^{**}(i', j', k') \\- I'(i', j', k')- Z_l)\bmod 256,
\end{multline*}
respectively.
\end{itemize}

\section{Chosen-plaintext attack}
\label{sec:cpa}

In \cite[Sec.~3.2.6]{WangXY:3operations:SP12}, it is claimed that the image encryption algorithm under study is robust against chosen-plaintext attack based on the following two points: (a) the used PRNSs are all sensitive to changes of secret key; (b) the substitution function~(\ref{eq:substitution}) owns a feed-back mechanism. However, we will show that the claim is not right in this section. As the image encryption algorithm under study is composed of three independent encryption operations, the position permutation part and the substitution part can be broken separately with a strategy of \textit{Divide and Conquer}.

As for plain-images of fixed value,
both the \textit{Row permutation} and the \textit{Column permutation} are canceled and only the \textit{Substitution} is left. Assume two chosen plain-images of fixed value $\bm{I}_1=\{I_1(i, j, k)\equiv d_1\}$, $\bm{I}_2=\{I_2(i, j, k)\equiv d_2\}$
are available. From Eq.~(\ref{eq:substitutionitem}), one has
\begin{equation}
 I'_1(0, 0, Y_0)=(I_1(0, 0, Y_0) + Z_0)\bmod 256
\label{eq:yequalsto0a}
\end{equation}
and
\begin{equation}
I'_2(0, 0, Y_0)=(I_2(0, 0, Y_0) + Z_0)\bmod 256.
\label{eq:yequalsto0b}
\end{equation}
Subtract Eq.~(\ref{eq:yequalsto0b}) from Eq.~(\ref{eq:yequalsto0a}), one has
\begin{equation}
(I'_1(0, 0, Y_0)-I'_2(0, 0, Y_0))\in\{D, D-256, D+256\},
\label{eq:possibleY0}
\end{equation}
where $D=d_1-d_2$. Referring to Eq.~(\ref{eq:substitution}), one has
\begin{IEEEeqnarray}{rCl}
I_1'(i, j, k) &=& (I_1(i, j, k) +I_1(i', j', k') \nonumber\\
                 &&+ \: I'_1(i', j', k')+ Z_{l})\bmod 256, \label{eq:substitutioncpa1}\\
I_2'(i, j, k) &=& (I_2(i, j, k) +I_2(i', j', k') \nonumber\\
                 &&+ \: I'_2(i', j', k')+ Z_{l})\bmod 256\label{eq:substitutioncpa2}
\end{IEEEeqnarray}
for $l=1\sim 3MN-1$, where $(i, j, k)$ and $(i', j', k')$ are determined by $\{Y_t\}_{t=0}^{l}$ and $\{Y_t\}_{t=0}^{l-1}$ respectively, as the above section.
Subtract Eq.~(\ref{eq:substitutioncpa2}) from Eq.~(\ref{eq:substitutioncpa1}), one has
\begin{equation}
(I'_1-I'_2)(i, j, k)\equiv (2D + (I'_1-I'_2)(i', j', k')) \pmod{256}
\label{eq:difference}
\end{equation}
where $(I'_1-I'_2)(i, j, k)=I'_1(i, j, k)-I'_2(i, j, k)$, and $(I'_1-I'_2)(i', j', k')=I'_1(i', j', k')-I'_2(i', j', k')$, the same hereinafter.

Then, a property of $(I'_1-I'_2)$ can be presented as follows.
\begin{Property}
Difference between the cipher-images of $\bm{I}_1$ and $\bm{I}_2$ satisfies that
\begin{equation}
(I'_1-I'_2)(i, j, Y_l) \equiv ((2l+1)D) \pmod{256}
\label{eq:property}
\end{equation}
for $l=0\sim 3MN-1$, where $(i, j)=(0, 0)$ when $l=0$, $(i, j)=(\lfloor (n_{Y_l}+1)/N \rfloor, (n_{Y_l}+1) \bmod N)$ otherwise, and
$n_{Y_l}$ denotes the number of the elements in $\{Y_t\}_{t=0}^{l-1}$, whose values are equal to $Y_l$.
\label{pro:aaa}
\end{Property}
\begin{proof}
This property can be proved via mathematical induction on $l$.
When $l=0$, one can get
\begin{equation*}
(I'_1-I'_2)(0, 0, Y_0) \equiv D \pmod{256}
\end{equation*}
from Eq.~(\ref{eq:possibleY0}), which means Eq.~(\ref{eq:property}) holds for $l=0$. Assume Eq.~(\ref{eq:property}) holds for $l=l^*$, i.e.,
\begin{equation*}
(I'_1-I'_2)(i, j, Y_{l^*}) \equiv ((2l^*+1)D) \pmod{256}
\end{equation*}
where $l^*<3MN-1$. Then, let us study the case for $l=(l^*+1)$.
From Eq.~(\ref{eq:difference}), one has
\begin{IEEEeqnarray}{rCl}
(I'_1-I'_2)(i, j, Y_{l^*+1}) &\equiv & (2D + (I'_1-I'_2)(i, j, Y_{l^*})) \pmod{256} \nonumber \\
                             &=      & ((2(l^*+1)+1)D) \pmod{256}. \nonumber
\end{IEEEeqnarray}
This completes the mathematical induction, hence finishes the proof of the property.
\end{proof}

Utilizing Property~1, one can get the estimated version of $Y_0$,
\begin{eqnarray}
\widehat{Y}_0=
\begin{cases}
0   &  \mbox{if } (I'_1-I'_2)(0, 0, 0) \equiv D \pmod{256},\\
1   &  \mbox{if } (I'_1-I'_2)(0, 0, 1) \equiv D \pmod{256},\\
2   &  \mbox{if } (I'_1-I'_2)(0, 0, 2) \equiv D \pmod{256},
\end{cases}
\end{eqnarray}
when $D\neq 128$. Obviously, one can assure $\widehat{Y}_0=Y_0$ definitely when
\begin{equation}
\#\left(\left\{k \;|\; (I'_1-I'_2)(0, 0, k) \equiv D \pmod{256}\right\}\right)=1,
\label{eq:condition1}
\end{equation}
where $\#(\cdot)$ denotes the cardinality of a set. Once the value of $Y_0$ is determined, the estimated values of
$\{Y_l\}_{l=1}^{3MN-1}$, $\{\widehat{Y}_l\}_{l=1}^{3MN-1}$, can be obtained in order with the similar method, namely
set
\begin{equation*}
\widehat{Y}_l=k \mbox{  \;if } (I'_1-I'_2)(i_k, j_k, k) \equiv ((2l+1)D) \pmod{256}
\end{equation*}
for $l=1\sim 3MN-1$, where $i_k=\lfloor (n_{k}+1)/N \rfloor$, $j_k=(n_{k}+1)\bmod N$, and
$n_{k}$ represents the number of $k$  in $\{\widehat{Y}_t\}_{t=0}^{l-1}$.

Referring to \cite[Sec. 5.4]{Hardy:Numberthoery:Oxford08}, one can get period of the sequence $\{(2l^*+1)D)\bmod 256\}_{l^*=0}^{3MN-1}$, $T=\frac{256}{2\cdot \gcd{(D,256)}}=\frac{128}{\gcd{(D,256)}}$. To help estimate success probability of this attack, we give another property of $(I'_1-I'_2)$ as follows.
\begin{Property}
Inequality
\begin{equation*}
\#\left(\left\{k \;|\; (I'_1-I'_2)(i_k, j_k, k) \equiv ((2l^*+1)D)  \pmod{256}\right\}\right)>1
\end{equation*}
holds if and only if
\begin{equation}
Y_{l^*+S}\not\in\{Y_l\}_{l=l^*}^{l^*+S-1},
\label{eq:errorcondition}
\end{equation}
where $({S}\bmod{T})=0$,
$i_k=\lfloor (n_{k}+1)/N \rfloor$, $j_k=(n_{k}+1)\bmod N$, and
$n_{k}$ represents the number of $k$  in $\{Y_t\}_{t=0}^{l^*-1}$.
\end{Property}
\begin{proof}
Assume, for the purpose of contradiction, that certain $l^*$ satisfies $Y_{l^*+S}\not\in\{Y_l\}_{l=l^*}^{l^*+S-1}$ and such that
\begin{equation*}
\#\left(\left\{k \;|\; (I'_1-I'_2)(i_k, j_k, k) \equiv ((2l^*+1)D)  \pmod{256}\right\}\right)=1.
\label{eq:condition3}
\end{equation*}
From Property~1 and the hypothesis, one has
\begin{equation*}
(2l^*+1)D \neq (2(l^*+S)+1)D \pmod{256},
\end{equation*}
which leads to
\begin{equation*}
0 \neq 2SD \pmod{256}.
\end{equation*}
Then, one has
\begin{IEEEeqnarray*}{rCl}
(S\bmod T) &=& S\bmod \frac{128}{\gcd{(D,256)}}\\
         &=& (2SD)\bmod \frac{2D\cdot 128}{\gcd{(D,256)}}\\
         &=&(2SD)\bmod \left(256\cdot \frac{D}{\gcd{(D,256)}}\right)\\
         &\neq & 0,
\end{IEEEeqnarray*}
thereby contradicting with the given condition. So, the property is proved.
\end{proof}

Assume that $Y_l$ uniformly distributes over $\{0, 1, 2\}$ for $l=0\sim 3MN-1$, one can
calculate the probability that condition~(\ref{eq:errorcondition}) in Property~2 hold for a given $l^*$ and $T$,
\begin{equation*}
Prob\left[Y_{l^*+S}\not\in\{Y_l\}_{l=l^*}^{l^*+S-1}\right]=\left(\left(\frac{2}{3}\right)^{kT}\cdot \frac{1}{3}\right),
\end{equation*}
where $k\in\{1,\cdots, \lfloor 2MN/T\rfloor\}$. Then, an upper bound of the probability that condition~(\ref{eq:errorcondition}) hold
can be got as
\begin{equation*}
Prob(MN)=\sum\nolimits_{k=1}^{\lfloor 2MN/T\rfloor}(3MN-kT)\left(\left(\frac{2}{3}\right)^{kT}\cdot \frac{1}{3}\right).
\end{equation*}
When $T=128$, one can calculate $Prob(2272\cdot 1704)\approx1.1173\cdot 10^{-16}$ for a relatively big plain-image of size
$2272\times 1704$. As for plain-images of smaller size, one can assure that the success probability of this attack is much bigger than
$(1-1.12\cdot 10^{-16})$ due to that the following points hold at the same time.
\begin{itemize}
\item The upper bound probability $Prob(MN)$ is a strictly increasing function with respect to $MN$;

\item Even Eq.~(\ref{eq:errorcondition}) holds, $\widehat{Y}_l^*=Y_l^*$ would still happen with probability $\frac{1}{2}$ or $\frac{1}{3}$;

\item The value of $Prob(MN)$ is calculated by summarizing the probability of some cases that may happen simultaneously.
\end{itemize}
Based on the above analysis, one can conclude that breaking of the \textit{Substitution} part can be implemented successfully with an extremely high probability.

Once the equivalent secret key determining \textit{Substitution} is recovered, the image encryption algorithm under study
becomes a position permutation-only gray-scale image encryption algorithm composing of the \textit{Row permutation} and the \textit{Column permutation}.
Considering the number of possible positions of every plain-pixel is $3MN$, the bit length of each element of chosen plaintext should be $\lceil\log_2(3MN)\rceil$ to assure that every permuted elements are different from each other. As bit size of every channel of plain-image is fixed to $8$, only $\lceil (\log_2(3MN))/8 \rceil$
pairs of chosen plain-images are required to recover the equivalent version of $\{T_l\}_{l=0}^{3M-1}$ and $\{T^*_{i,l}\}_{i=0,l=0}^{M-1,3N-1}$.
Referring to quantitative cryptanalysis of permutation-only encryption algorithms in \cite{Li:AttackingPOMC2008,Lcq:Optimal:SP11}, the complexity of breaking the position permutation part is only $O(3MN)$.

\begin{figure}[!htb]
\centering
\begin{minipage}[t]{\imagewidth}
\centering
\includegraphics[width=\imagewidth]{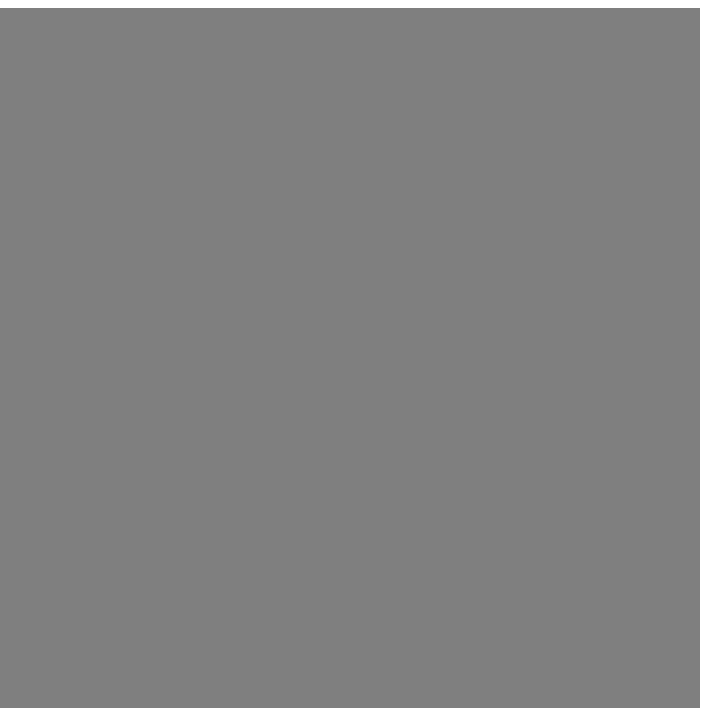}
a)
\end{minipage}
\begin{minipage}[t]{\imagewidth}
\centering
\includegraphics[width=\imagewidth]{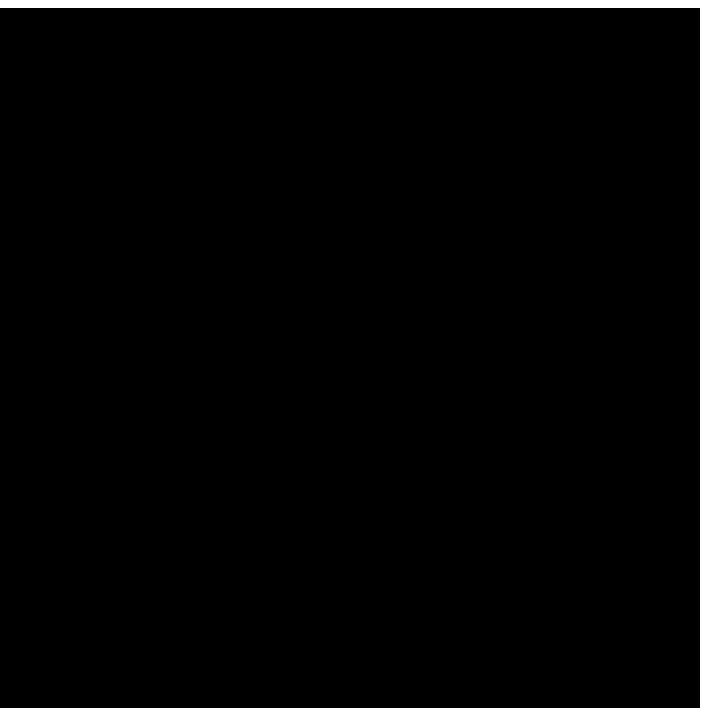}
b)
\end{minipage}
\\
\begin{minipage}[t]{\imagewidth}
\centering
\includegraphics[width=\imagewidth]{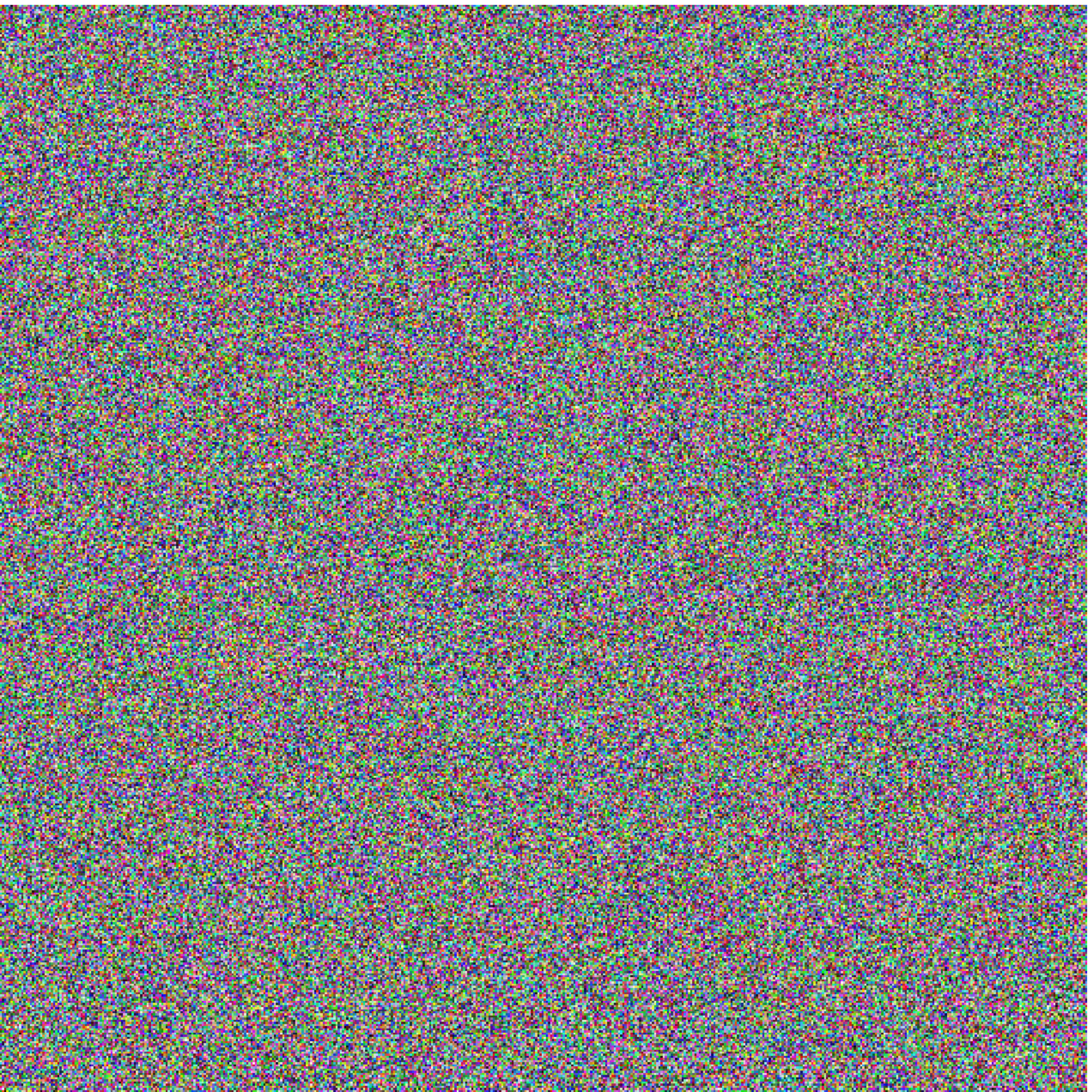}
c)
\end{minipage}
\begin{minipage}[t]{\imagewidth}
\centering
\includegraphics[width=\imagewidth]{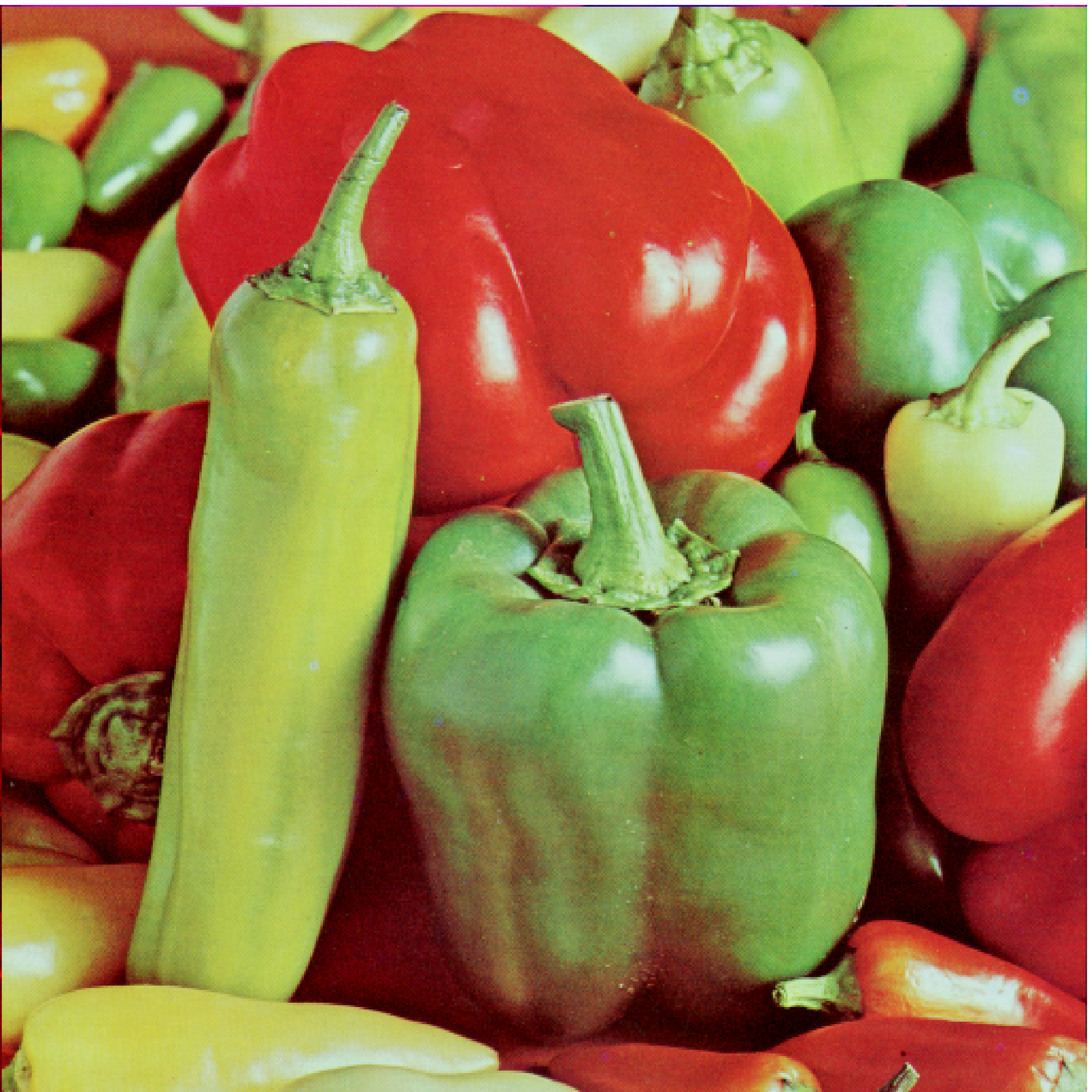}
d)
\end{minipage}
\caption{Chosen-plaintext attack:
a) the chosen plain-image of fixed value $127$;
b) the chosen plain-image of fixed value $0$;
c) the cipher-image of plain-image ``Baboon";
d) the recovered plain-image of the image shown in Fig.~\ref{figure:decryption2}c).}
\label{figure:decryption2}
\end{figure}

To validate the performance of the proposed attack, a great number of experiments on some plain-images of size $512\times 512$ were made
with some randomly selected secret keys. When $\mu_0=4.0$, $x_0=0.123456789764$, $m_1=1000$, $\mu_0^*=3.999999$, $x^*_0=0.567891234567$ and $m_2=2000$,
two chosen plain-images of fixed pixel value $127$ and $0$, shown in Fig.~\ref{figure:decryption2}a) and b) respectively, are used to recover the PRNS $\{Y_l\}_{l=0}^{3MN-1}$. Then, $\lceil (\log_2(3\times 2^9\cdot 2^9))/8 \rceil=3$ pairs of chosen plain-image are constructed to recover the equivalent secret of the position permutation-only part.
Finally, the equivalent versions of the sub-keys controlling two main encryption parts are used together to break a cipher-image encrypted with the same secret key, which is shown in Fig.~\ref{figure:decryption2}c). The decryption result is shown in Fig.~\ref{figure:decryption2}d) and it is identical with the original plain-image, which verifies the effectiveness of the proposed attack.

\section{Conclusion}

This paper studied the security of a novel colour image encryption algorithm based on chaos proposed in \cite{WangXY:3operations:SP12}.
It is found that the encryption algorithm can be broken with chosen-plaintext attack efficiently. The number of required chosen plain-images and complexity
of the attacking are proportional to a logarithm of size of plain-images and the size, respectively. As a conclusion, the image encryption algorithm under study is not suggested in serious applications requiring a high level of security.

\section*{Acknowledgement}

This research was supported by the National Natural Science Foundation of China (No.~61100216), Scientific Research Fund of Hunan Provincial Education Department (No.~11B124),
and Start-up Fund of Xiangtan University (Nos.~10QDZ39,~10QDZ40).

\bibliographystyle{elsarticle-num}
\bibliography{SP12}
\end{document}